# Pulse tube cryocooler with self-cancellation of cold stage vibration

Suzuki T., Tomaru T., Haruyama T., Shintomi* T., Sato N., Yamamoto A.,
Ikushima Y **., Li R. **

High Energy Accelerator Research Organization, Tsukuba-shi, Ibaraki-ken 305-0801 , Japan
*Advanced Research Institute for the Sciences and Humanities Nihon University,
   Chiyoda-ku, Tokyo-to 102-0073, Japan
** Sumitomo Heavy Industry Inc., Nishitokyo-shi, Tokyo-to 188-8585, Japan

We experimentally demonstrate a new method for reducing the vibration of cold stage of a cryocooler. A model cryocooler consists of two pairs of a pulse-tube and a regenerator with identical structures. Each pair was connected with a common cold stage. Comparing RMS amplitude with the case of no phase shift of driving gas pressure between the two pairs, the longitudinal vibration of the cold stage reduces 96.1% at 126K by supplying the gas pressure with 180 degrees of phase shift.

INTRODUCTION

Cryocoolers are used for cooling many kinds of cryogenic instruments because of their convenience and compactness. Particularly the pulse tube cryocooler is expected to apply to the instruments or the devices that are easily interfered with external vibration. A typical value of the cold stage vibration of an ordinary pulse tube cryocooler is about ten micrometers, although the vibration of cold head is about 100 times smaller than that of the Gifford-McMahon cryocooler[1]. Even with the pulse tube cryocooler, vibration of the cold stage is still destructive for applications to extremely sensitive instruments such as gravitational wave detectors. So it has been required to develop an advanced pulse tube cryocooler with smaller vibration. Current solution is to put an additional vibration isolator on the pulse tube cryocooler. In this situation, we proposed the method of self-cancellation for reducing the vibration of the cold stage[2]. The method of self-cancellation can keep convenience and compactness of the pulse tube cryocooler.

   The vibration of the cold head results from a driving mechanism of the pulse tube cryocooler. An alternating pressure of gas, which is generated by a compressor and an appropriate valve mechanism, drives the pulse tube cryocooler. The variation of the pressure causes a vibration of the cold stage through elastic deformation of pipes of the regenerator and the pulse tube. The vibration of the cold stage can be considered as a forced oscillation by the alternating pressure of gas. The driving gas pressure works as an external force that induces the cold stage vibration. One of the ideas to reduce the cold stage vibration is to use the gas pressure itself as a counterforce. The motion of the cold stage keeps stable when the resultant of external forces is zero.

   A simple example for realizing this is to connect a pair of pulse tube cryocoolers on the same cold stage and supply the periodic gas pressure with 180 degrees phase shift to each cryocooler. The vibrations of the cold stage will be cancelled out when a configuration of the cryocoolers has a proper symmetry on the cold stage. The configuration can be generalized to the case of n-pair cryocoolers with the same cold stage. A preliminary experiment showed that the cold stage vibration reduced more than 96% at 300K[2].

In order to prove the method of self-cancellation, we measured vibrations of the cold stage of a model cryocooler. The model cryocooler is essentially a complex of two pulse tube cryocoolers with identical structures. In the followings, details of the experiment will be described.

EXPERIMENTAL PROOF FOR THE METHOD OF SELF-CANCELLATION

Experimental setup
An experimental setup is shown in Figure 1. The model cryocooler was installed in the vacuum chamber. A valve unit generates an alternating gas pressure by switching high and low pressures supplied from a compressor. Gas pressures on each cryocooler were monitored by pressure transducers (PT). The valve timing controller supplied control pulses to the valve unit. Pulses with a period about 0.5 sec were generated by the circuit of timing IC. Vibration of the cold stage was measured by the optical transducer. The optical transducer was fixed on the table which is rigidly connected to the top flange of the vacuum chamber. Two silicon diode thermometers monitor temperatures on the model cryocooler. Buffer volumes were connected to pulse tubes by needle valves (NV). All of the data were stored in the data logger.

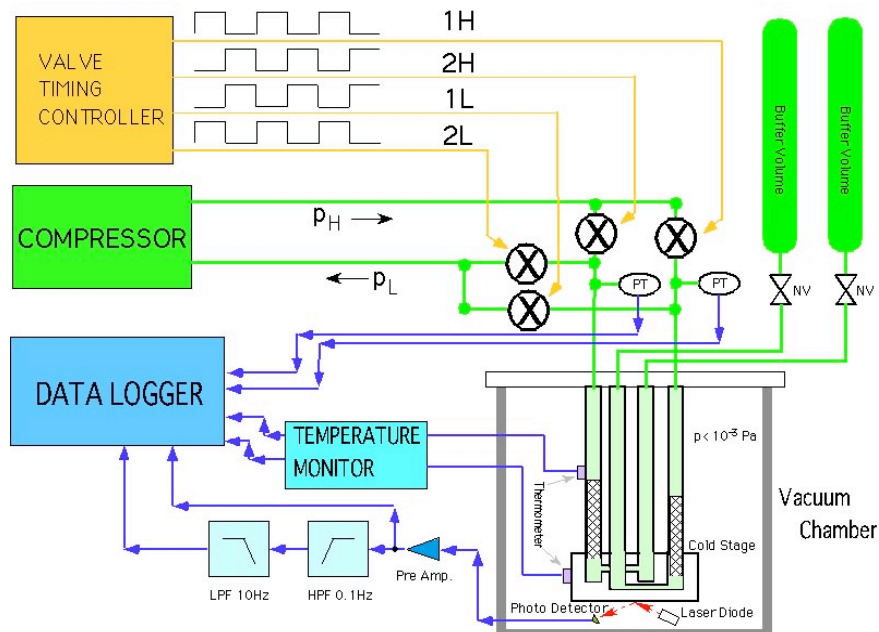

Figure 1 Setup of the experiment.

The model cryocooler
Four stainless-steel tubes, which were two regenerators and two pulse tubes, were joined on the cold stage. Tubes are separated in 90 degrees on the same circle. Each two tubes which held the common diameter were connected by a horizontal hole in the cold stage to make a cryocooler unit. Sizes of the tubes are 20mm of diameter, 300mm of length and 0.25mm of thickness. Figure 2 shows the model cryocooler.

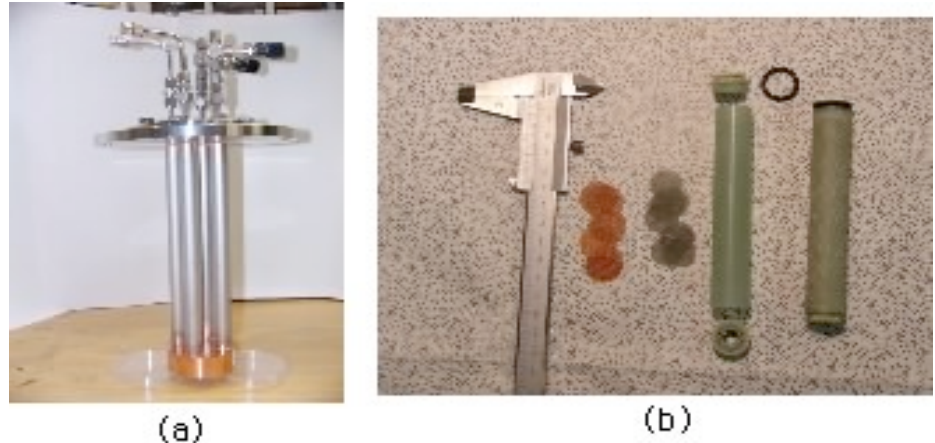

Figure 2 The model cryocooler (a) and the cartridges of regenerator.

The regenerator cartridge consists of meshes (#100, stainless-steel) and a pipe with caps (G10) at both ends. The stainless-steel meshes were packed into the G10 pipe to a mass about 59.1g to 61.1g. The caps were joined to the pipe by a drop of cyanoacrylate glue. An O-ring at one end of the cap stops leakage between the cartridge and the regenerator tube. Copper meshes were used as spacers in the bottom of regenerator tubes.

The optical transducer
The optical transducer consists of a semiconductor laser module and a photo detector (PD). The PD monitors the light that was reflected 45 degrees by a small mirror on the bottom of the cold stage. The amount of an incident light to the PD was modulated by a distance between the PD and the mirror when the position of the PD to the light beam had an appropriate offset. The optical transducer was fixed on the translation stage that could control the distance between the PD and the mirror. The conversion coefficient of the optical transducer was 59.84 micrometers/V when the DC output was about 3 volts.

Operation of the model cryocooler
The supply and the return pressures of the compressor were respectively 1.45 MPa and 0.85 MPa. Both the compressor and the valve unit were located in the neighbor room to avoid their mechanical vibration and acoustic noise. Copper tubes with 1/4 inch of diameter connected the valve unit and the model cryocooler. The volume of buffer tanks was approximately 0.6 liter. Needle valves were adjusted to an appropriate opening extent to be able to achieve the minimum cold head temperature. The optical transducer was set such that the DC component of the transducer showed to be about 3 volts.

RESULTS

The temperature of the cold head reached to 126 K about 90 minutes after the start of operation. In the same time, the middle of the regenerator tube was almost at the room temperature. It was enough for elastomer of the O-rings to keep soft. Phase shifts of supplied gas pressures were set to be two values. The 0 degrees phase shift corresponded to an ordinary operation of cryocooler and the 180 degrees phase shift corresponded to applying the self-cancellation operation. The RMS (Root Mean Square) amplitude was 2.94 micrometers for the case of 0 degrees phase shift and was 0.115 micrometers for the case of 180 degrees phase shift. Also the noise level of the measurement was 0.033 micrometers in

RMS. Figure 3 shows the results of vibration measurement at 126 K.

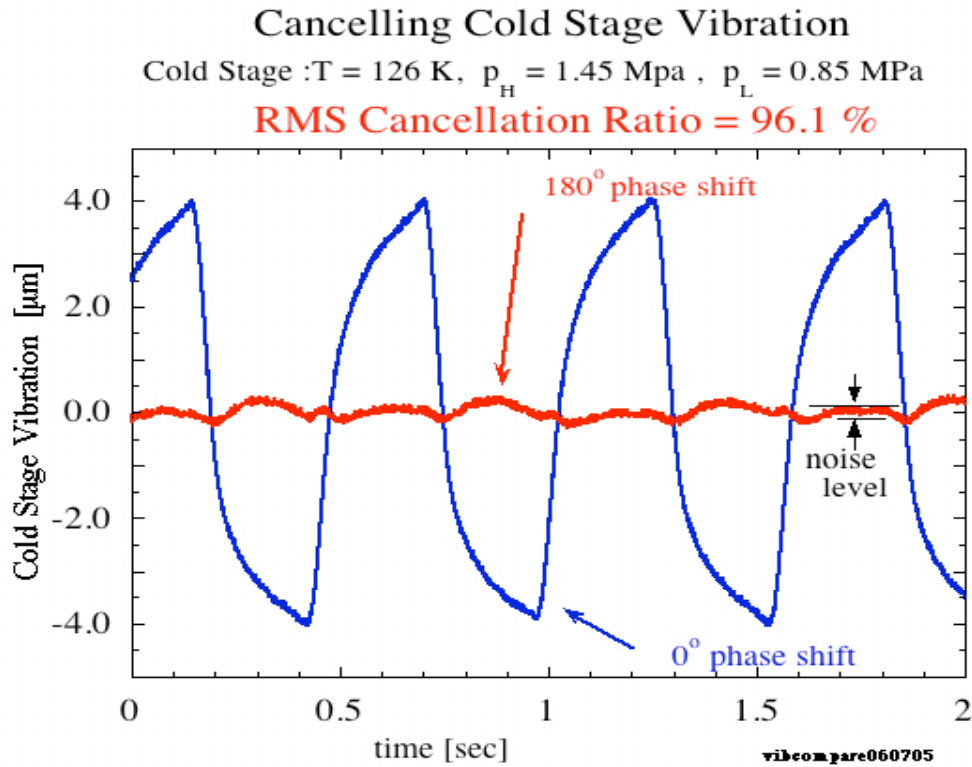

Figure 3 The vibration of the cold stage of the model cryocooler. The blue line shows the operation in 0 degree phase shift. The red line shows the operation in 180 degrees phase shift. The width of the red line corresponds to the noise level.

CONCLUSION

Comparing the RMS amplitudes, the vertical vibration of the cold stage reduced 96.1% by applying the self-cancellation from the ordinary pulse tube cryocooler. From those results, we conclude that the method of self-cancellation is effective to reduce the vibration of the cold stage.

ACKNOWLEDGMENT

This work was supported by the Grant-in-Aid for Scientific Research 16360110. We thank to Mr. K.Kasami and Mr. N.Kudo for their assistance in fabricating apparatus.